\documentstyle[11pt,aaspp4]{article}

\lefthead{Rao et al.}
\righthead{Magnetic Fields in Orion}

\begin{document}
\singlespace

\title{HIGH RESOLUTION MILLIMETER-WAVE MAPPING OF LINEARLY POLARIZED
DUST EMISSION: MAGNETIC FIELD STRUCTURE IN ORION}

\author{R. Rao\altaffilmark{1} and R. M. Crutcher}
\affil{Department of Astronomy, University of Illinois,
    Urbana-Champaign, IL 61801}

\and
\author{R. L. Plambeck and M. C. H. Wright}
\affil{Radio Astronomy Laboratory, University of California,
    Berkeley, CA 94720}

\altaffiltext{1}{email: ramp@astro.uiuc.edu}

\begin{abstract}

We present 1.3 and 3.3 mm polarization maps of Orion-KL obtained with
the BIMA array at approximately $4\arcsec$ resolution.  Thermal emission from
magnetically aligned dust grains produces the polarization.  
Along the Orion ``ridge'' the polarization position angle varies smoothly from
about $10\arcdeg$ to $40\arcdeg$, in agreement with previous lower
resolution maps.  In a small region south of the Orion ``hot core,''
however, the position angle changes by $90\arcdeg$.  This abrupt change
in polarization direction is not necessarily the signpost of a twisted
magnetic field.  Rather, in this localized region processes other than
the usual Davis-Greenstein mechanism might align the dust grains with
their long axes parallel with the field, orthogonal to their normal
orientation. 

\end{abstract}


\clearpage

\section{INTRODUCTION}

Magnetic fields play many roles in the star formation activity in
molecular clouds (see review by \cite{mck93}).
Polarization of radiation is one of the important signatures of
interstellar magnetic fields (\cite{hil88}).  
Spinning dust grains in the interstellar
medium become partially aligned with the magnetic field, generally with
their long axes perpendicular to the field (\cite{dav51}; hereafter DG). 
Thus, the thermal emission from grains at far infrared and millimeter
wavelengths is partially linearly polarized, with polarization direction
perpendicular to the magnetic field. 

As the nearest region of OB star formation, the Orion molecular cloud
has been studied intensively (see review by \cite{gen89}).  In the
vicinity of the Kleinmann-Low Nebula (KL) the cloud contains at least
two massive stars, IRc2 and the Becklin-Neugebauer object (BN), embedded
within a flattened ``ridge'' of molecular gas which extends along a
position angle of 30$\arcdeg$.  Polarization from aligned dust grains in
Orion has been mapped at 100 $\micron$ with 35--40$\arcsec$ resolution
(\cite{nov89}; \cite{gon90}; \cite{sch98}), at 350 $\micron$ and 450 $\micron$
 with 18$\arcsec$ resolution (\cite{sch97}; \cite{sch98}), at 800 $\micron$ with
14$\arcsec$ resolution (\cite{ait97}), and at 1.3 mm with 30$\arcsec$
resolution (\cite{lea91}).  In all these maps the polarization vectors
are roughly parallel with the molecular ridge, indicating that the large
scale magnetic field is perpendicular to the ridge.  The uniformity of
the polarization direction across the region suggests that the field is
quite strong, of order 1 mG (\cite{gon90}; \cite{lea91}).  The
fractional polarization ranges from 4\% to 8\% except along the line of
sight directly through KL, where it is significantly lower. 

Higher angular resolution is needed to probe the polarization pattern
near KL.  Heretofore, this has been possible only at near- and mid-IR
wavelengths, where the competing effects of absorption, emission, and
scattering all influence the polarization direction.  Some of the most
secure results are obtained by mapping the polarization of the 2~$\micron$
 S(1) line of H$_2$ (\cite{hou86}; \cite{bur91}; \cite{chr94}).  The
line emission (assumed unpolarized)  originates from
shock-excited H$_2$ in the bipolar outflow from IRc2.  Absorption by
aligned grains in front of the outflow produces the polarization. 
Aitken et al.\ (1997) also mapped the continuum polarization at 12.5 and
17 $\micron$; spectropolarimetry suggests that it, too, is mostly
attributable to absorption.  These studies all find that the
polarization vectors are twisted near KL.  The authors argue that this
is evidence for a toroidal magnetic field in a disk-like structure
centered on IRc2. 

In this paper we present the first interferometric polarization maps of
Orion at millimeter wavelengths, where the polarization arises
unambiguously from emitting grains.  These high resolution maps confirm
the abrupt change in polarization direction near IRc2 first detected in
the 2 $\micron$ data.  We argue that in this small region the bipolar
outflow from IRc2 might align the grains with their long axes parallel
with the magnetic field, orthogonal to their usual orientation.  Thus,
it is not certain that the change in the polarization direction arises
from a twisted magnetic field.

\section{OBSERVATIONS AND DATA ANALYSIS}

Observations were made with the 10-antenna BIMA array
(\cite{wel96}) in 1997 March,
November, December and 1998 January.  Data were obtained at 90 GHz
($\lambda$ 3.3 mm) with the antennas in the B and C configurations, and
at 230 GHz ($\lambda$ 1.3 mm) with only the C configuration.  

Linear polarization measurements with interferometers are best done with
circularly polarized feeds (see \cite{tho86}).  This minimizes the
effect of gain errors, as the cross-correlation of opposite circular
polarizations does not involve the Stokes total intensity (I) parameter. 
Quarter wave plates are used in front of the linearly polarized BIMA 
feeds to obtain left (L) or right (R)
circular polarizations.  Since only a single polarization is received, L
and R are time-multiplexed on each antenna using a fast Walsh function
switching pattern in order to sample all possible cross correlations
(LL, LR, RL, RR) on every baseline.  
The data were averaged over the Walsh cycle to produce
quasi-simultaneous dual polarization measurements.
Beam smearing resulting from this averaging process is negligible.

The instrumental polarization response, or ``leakage,'' for each
antenna was calibrated by observing a strong point source (3c273 or
3c279) over a wide hour angle range
to provide good parallactic angle coverage.  For antennas with alt-az
mounts and orthogonal circular feeds, the fringe phases for a linearly
polarized source vary with parallactic angle while the instrumental
leakages remain constant, so one can solve simultaneously for the
leakages and the source polarization (\cite{sau96}).
Typical leakage amplitudes were 1\% at 3.3 mm and 5\% at 1.3
mm. Leakages measured on different days or on different
calibrators were consistent within 0.3\% rms at 3.3 mm and 0.5\% rms at 1.3 mm. 
Test observations also showed that the instrumental polarization does not
vary strongly across the primary beam.  
The QSO 3c286 was used to check the position angle calibration.

The data were reduced with the MIRIAD software package (\cite{wri93};
\hbox{http://www.atnf.csiro.au/computing/software/miriad}).
The task GPCAL was used to fit the instrumental leakage terms on the
polarization calibrator.  The averaged 
Orion data were corrected for the leakages,
and a 400 MHz wide continuum band without obvious spectral line contamination
was used to produce maps of the I, Q, U, and V Stokes parameters.
These maps were deconvolved independently and then 
combined to produce maps of the linear polarization intensity, the
fractional polarization and the 
position angle.  The synthesized beam size was 6.8$\arcsec\times$2.8$\arcsec$ 
at 3.3 mm and 4.4$\arcsec\times$2.4$\arcsec$ at 1.3 mm.

\section{RESULTS}

\placefigure{fig1}
\placefigure{fig2}

Figures 1 and 2 display the 3.3 mm and 1.3 mm polarization maps.  Almost
all of the emission at these wavelengths is thermal radiation from dust. 
We have not corrected the maps for contamination by free-free emission
from the compact circumstellar envelopes associated with BN and IRc2 (30
to 150 mJy; \cite{pla95}), nor from the extended foreground \ion{H}{2}
region (important only along the eastern edge of the 3.3 mm map, judging
from Figures 1 and 2 of \cite{wri92}). 

Polarization vectors are plotted only where the linearly polarized
intensity $\sqrt{Q^2 + U^2}$ is greater than 3 $\sigma$, where $\sigma$ 
(the rms noise level in the Q and U maps)
is 1.8 mJy/beam at 3.3 mm and 12 mJy/beam at 1.3
mm.  This cutoff corresponds to an uncertainty of approximately
10$\arcdeg$ in the polarization position angle PA = $\frac{1}{2}\tan^{-1}
\frac{U}{Q}$.  The percentage polarizations we observe are as high as
20$\%$ in some directions, particularly near the edges of the emission
region.  Such high values probably are not significant.  They can occur
if the Stokes I emission region is more extended than the linearly
polarized region.  In that case the interferometer resolves out some of
the extended emission in the I-image, but detects most of the linearly
polarized flux, so one overestimates the fractional polarization.  
Fortunately, we are
interested principally in the polarization direction, which is unlikely
to be corrupted by this effect.  

\placetable{tbl-1}

Table~\ref{tbl-1} lists the polarizations and formal errors for 4
independent positions (separated by at least one synthesized beamwidth) 
where we have both 3.3 mm and 1.3 mm data.  
The 1.3 mm maps
were convolved to the same resolution as the 3.3 mm maps for this
comparison.  The 1.3 mm and 3.3 mm polarization position angles
differ by at most 15$\arcdeg$, which is
within the expected uncertainties.  
The rotation
angle due to Faraday rotation is given by $\Omega={\lambda}^2$ RM, where
$\lambda$ is the wavelength and RM is the rotation measure.  From
measurements at two different frequencies, one can compute the rotation
measure as RM$=$ $\Delta \Omega$/(${\lambda_2}^2 - {\lambda_1}^2$). 
Our data imply $\Delta \Omega < $ 0.25 radians, so RM $< 2.8 \times
10^4$ rad m$^{-2}$.  Using this upper limit for the rotation measure, we
find $\int n_e \, B_{\parallel} dl < 0.035$ G cm$^{-3}$ pc.
Thus, if the foreground
\ion{H}{2} region has an electron density of $10^3$ cm$^{-3}$ over a
distance of 1 parsec (\cite{wil87}), the average  line of sight
magnetic field through the \ion{H}{2} region is $\lesssim 35\, \mu$G.

\section{DISCUSSION}

The most striking feature in our maps is the abrupt change in
polarization direction $\sim$5$\arcsec$ south of IRc2.  The polarization
position angle rotates from 0$\arcdeg$ to 90$\arcdeg$ as one moves
eastward from the Orion ridge through this ``anomalous'' region, on both
the 1.3 mm and the 3.3 mm maps.  The twisted polarizations south of IRc2
were not detected in previous mm, submm, and FIR polarization maps
because of inadequate angular resolution.  When convolved to 20$\arcsec$
resolution , our 3.3 mm polarization map closely resembles the
earlier maps (cf.  \cite{lea91}; \cite{sch98}).  In particular, it shows
the polarization ``hole'' toward KL which results from averaging
across the region where the polarization direction varies.

Our data are consistent with the 2 $\micron$ polarization maps of the S(1)
line of H$_2$ (\cite{hou86}; \cite{bur91}; \cite{chr94}).  At 2~$\micron$
the polarization is produced by dust absorption, rather than emission,
so the polarization vectors should be orthogonal to the ones we
measure.  If we overlay our data on the 2 $\micron$ map shown in Figure 1
of Chrysostomou et al.\ (1994), the polarization vectors are orthogonal
almost everywhere, including the anomalous region.  Comparing our data
with the 12.5 $\micron$ and 17 $\micron$ polarization maps of Aitken et al. 
(1997) is more difficult because the mid-IR brightnesses are low
southeast of IRc2.  These maps do show, however, that southwest of IRc2
the polarization direction is roughly orthogonal to that along the Orion
ridge, similar to our results.

The anomalous polarization region lies along the southern edge of the
Orion ``hot core'' dust continuum peak.  Molecular line observations
(\cite{bla87}; \cite{gen89}; \cite{wri96}) indicate that gas kinetic
temperatures in the hot core range from 150 to 300 K, with densities
n(H$_2$) $\sim 10^7 \, \rm{cm}^{-3}$.  The Orion ``compact ridge''
feature overlaps the hot core along this same line of sight; it is a bit
cooler (80-140 K) and less dense ($10^6 \, \rm{cm}^{-3}$).  
It is likely that the hot core is physically near IRc2, probably within
1000 AU.  The compact ridge also
seems to be closely linked to the hot core and IRc2.  By contrast, the
ridge emission west of the hot core, where the polarization position
angles are ``normal'' probably arises in gas farther from IRc2. 

Many possible mechanisms can align dust grains in a molecular cloud,
and hence produce linear polarization (\cite{laz97c}, Table 1).  
Before discussing these processes, it is worthwhile to emphasize that,
regardless of the
alignment mechanism, the polarization direction must {\it always} be
correlated with the magnetic field direction in the cloud (\cite{mar71};
\cite{rob96}).  This is because a spinning grain acquires a magnetic
moment, primarily through the Barnett effect, and the interaction of
this moment with the magnetic field causes the grain to precess around
the field direction.  For a 2000 \AA\ diameter grain in a 1 mG field,
this Larmor precession period is only a few hours (\cite{rob96}), orders
of magnitude shorter than the time required for any alignment mechanism
to change the grain's angular momentum.  Thus, the net polarization from
an ensemble of grains, all precessing around the field, must be
symmetrical with respect to the magnetic field, either {\it parallel} or
{\it perpendicular} to it.  

In most regions of the interstellar medium grains are assumed to be
aligned via paramagnetic relaxation (the DG mechanism), which damps the
components of a grain's angular momentum perpendicular to {\bf B},
gradually aligning the spin axis with the magnetic field direction. 
Internal damping due to Barnett relaxation normally causes the grain to
spin about its axis of greatest moment of inertia (\cite{pur79}), so the
long axis of the grain is oriented perpendicular to the field. 
Presumably this is the situation along most of the Orion ridge.  If the
grains in the anomalous region also are aligned via the DG mechanism,
then either the magnetic field in this region must be twisted or 
these grains are rotating about their axes of least moment
of inertia.
The latter  occurs where the dust grains are hotter than the gas
(\cite{jon67}). There is little indication that this is the case in
Orion: the 20--30 $\micron$ dust color temperature across the entire KL
region is roughly 100 K (\cite{wyn84}), less than or comparable to the
gas kinetic temperatures inferred for the hot core and compact ridge. 
One may also dismiss this possibility 
because it leads to extremely low alignment efficiencies (\cite{laz97c}, Figure
2).  
We conclude that, if DG alignment applies, the magnetic
field in the anomalous region must be almost orthogonal to that in
the ridge. 

Chrysostomou et al.  (1994) and Aitken et al. 
(1997) have modeled this as the result of a toroidal magnetic field in
a dense circumstellar disk associated with IRc2. 
We are suspicious of this hypothesis for several reasons.  First, the
anomalous polarization region is not centered on IRc2.  Second, in our
maps the polarization direction does not seem to twist smoothly, but
instead jumps abruptly by 90$\arcdeg$, at the edge of the anomalous
region.  Finally, in a region as dense as the hot core it is difficult
for DG alignment to overcome the randomizing effect of gas-grain
collisions. The success of the DG mechanism
would require suprathermal rotation (\cite{pur79})  and/or
superparamagnetic damping (\cite{jon67}).

Although DG alignment in the hot core cannot be ruled out, we are
motivated to consider alternative alignment mechanisms which might
produce the observed change in polarization direction without invoking a
twisted magnetic field.  One such mechanism is alignment by the
intrinsic angular momentum of photons (\cite{har70}).  In an
anisotropic radiation field, absorbed photons preferentially excite the
component of a grain's angular momentum parallel to the photon flux, so the grain tends
to rotate with its long axis perpendicular to the photon flux.
Purcell \& Spitzer (1971)
showed that cold grains cannot be aligned by an anisotropic flux of optical
photons because roughly 300 infrared photons, each with intrinsic angular
momentum $\pm\hbar$, are emitted in random directions for each optical
photon absorbed.  Aitken et al. (1985) pointed out, however, that the mechanism
could be important for small grains close to a cool source of high
luminosity (such as IRc2),
where infrared, not optical, photons cause the alignment.  
However, the calculations given 
in $\S 4.1$ of Aitken et al. (1985) show that in a
region as dense as the Orion hot core
gas-grain collisions impart 10 times the mean squared angular
momentum of infrared photons (using $n\sim10^6 \, {\rm
cm}^{-3},\, a\sim10^{-5}\, {\rm cm},\, T_{gas} \sim 150$ K), so the
alignment efficiency is very low.

A more promising mechanism is Gold (1952) alignment, in which gas streaming
past the grains excites their angular momenta perpendicular to the flow 
direction, causing the grains to rotate with their long axes parallel to 
the flow.  Efficient alignment takes place if the flow speed is greater than
the random thermal velocities of the gas particles. 
IRc2 is known to be the source of a supersonic
bipolar outflow which certainly is adequate to align grains.
The outflow is well traced by
the CO millimeter transitions and is roughly perpendicular to the ridge,
at an angle of 15$\arcdeg$ to the large scale magnetic field
direction.  Lazarian (1994, 1997) modified the Gold alignment theory 
to take into account the
effects of suprathermal rotation, Larmor precession, and internal
alignment via Barnett relaxation.  His theory predicts that oblate
grains will be aligned with their long axes parallel with the magnetic
field if the angle between the outflow and the field is less than
${\rm \cos}^{-1}[1/\sqrt{3}]$ ($\theta\sim 55\arcdeg$). The alignment 
efficiency decreases as the angle between the outflow and the magnetic field
increases until this critical angle is reached.
If Gold alignment is responsible for the anomalous polarizations south of
the hot core, then the magnetic field throughout the entire Orion-KL region
could be relatively straight.

\section{CONCLUSIONS}

We have made the first interferometric observations of the polarization
from dust emission in the Orion BN/KL region, at wavelengths of both 1.3
mm and 3.3 mm.  We find that the decrease in fractional polarization
toward KL previously seen with single dish observations is due to
polarization structure which is averaged out by larger beams.  In
particular, our maps show that the polarization direction changes
abruptly by 90$\arcdeg$ in a small region south of the Orion hot core. 
If the grains are aligned everywhere by the Davis-Greenstein mechanism,
then the magnetic field in this anomalous region is almost orthogonal to
the large scale field in the Orion ridge.  However, it is plausible that
the grains in this region are aligned by a wind from IRc2, and that the
magnetic field is relatively straight.  These results suggest that one
should be cautious in using polarization data to infer the magnetic
field structure around young stars, because of the possibility that the
grains are mechanically aligned by outflows.

\acknowledgements

We are grateful to J. B. Lugten, G. Engargiola, D. Zhou, J. R. Forster
and the staff at Hat Creek Radio Observatory
for assistance in various stages
of the project. We also thank  A. Lazarian for helpful
discussions on grain alignment. This work 
was partially supported by NSF grants AST
94-19227, AST 96-13998, and AST 96-13999, and by the University of Illinois
and the University of California.

\clearpage

\figcaption[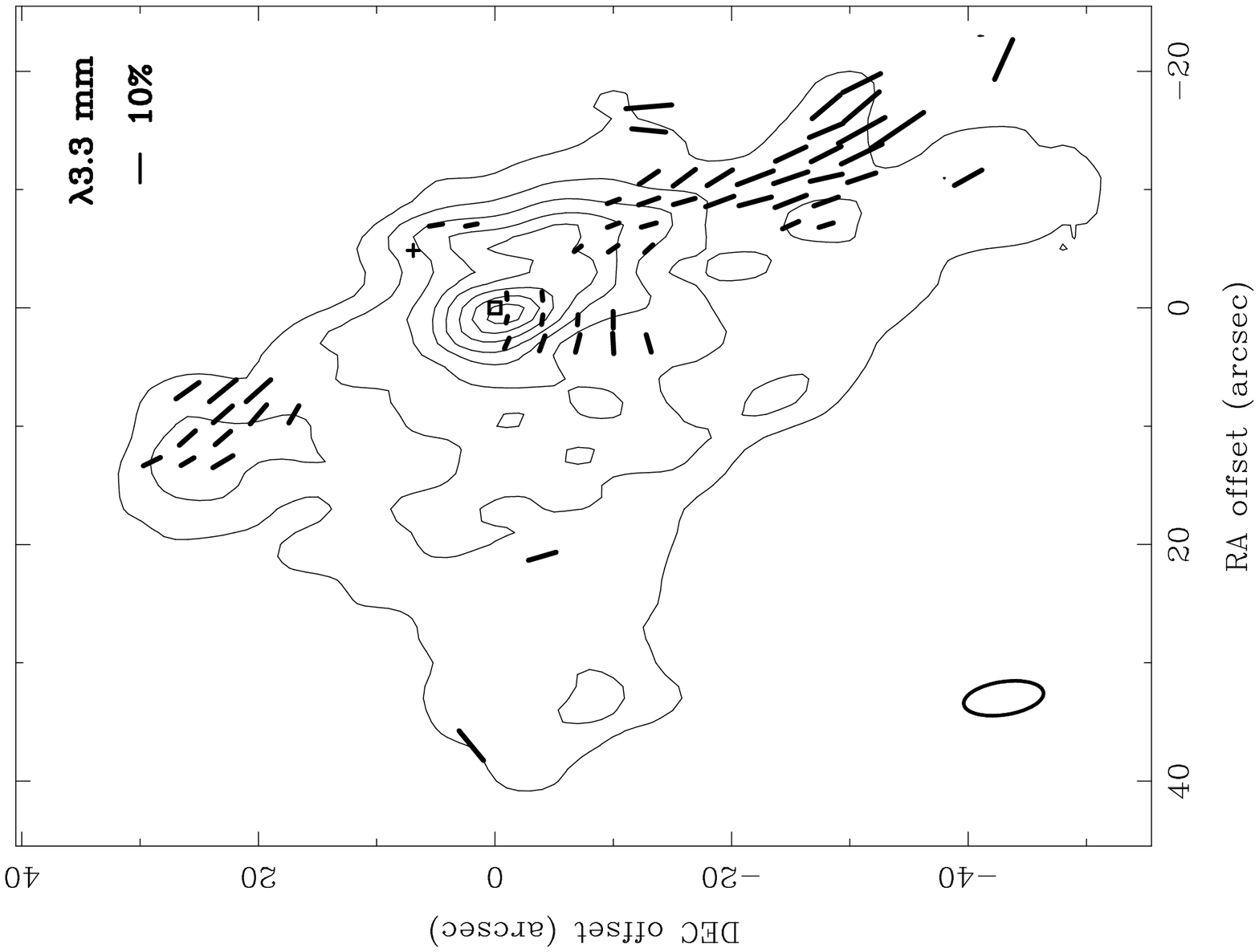]{ Polarization map of Orion-KL at 3.3 mm.  
Contours indicate the total 3.3 mm flux density (Stokes I) and are drawn at 12, 25,
38, 50, 60, 72, 84, 95 percent of the peak value of 360 mJy/beam.  The
noise level was $\sim$15 mJy/beam.  Vectors show the 
percentage polarization and position angle wherever linearly polarized flux was
detected with 3$\sigma$ or greater significance. 
Offsets are relative to IRc2--source I
at $\alpha = 5^h35^m14^s.505$, $\delta = -5\arcdeg
22\arcmin 30\arcsec .45$ (J2000).  The position of IRc2 is indicated by
a hollow square, BN by a cross.
The 6.8$\arcsec \times 2.8\arcsec$ synthesized beam is shown by the ellipse
at lower left.
\label{fig1}}

\figcaption[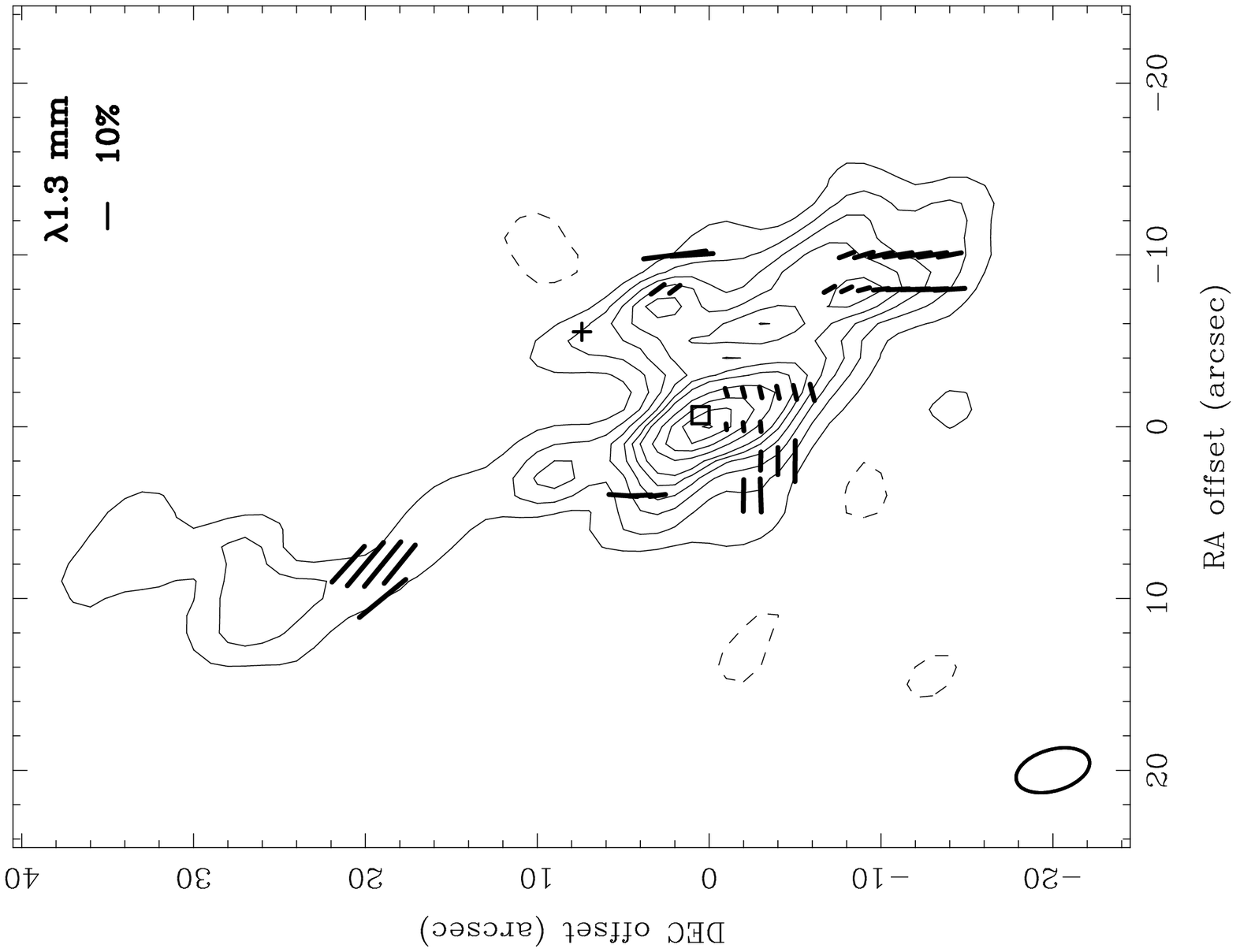]{ Polarization map of Orion-KL at 1.3 mm.  
Contours indicate the total 1.3 mm flux density and are drawn at -7, 7,
14, 21, 28, 35, 42, 56, 70, 84, 98 percent of the peak of 2.4 Jy/beam. 
The noise level was $\sim$46 mJy/beam.  Vectors and beam representation
as in Figure 1.  The synthesized beam is
4.4$\arcsec \times 2.4\arcsec$.
\label{fig2}}

\clearpage
 
\begin{deluxetable}{lccccc}
\footnotesize
\tablecaption{COMPARISON OF POLARIZATIONS AT $\lambda$3.3 MM AND $\lambda$1.3
MM \label{tbl-1}}
\tablewidth{0pt}
\tablehead{
\colhead{Region} & \colhead{Offset Position\tablenotemark{a}} & \colhead{PA at 1.3 mm} &
\colhead{PA at 3.3 mm} & \colhead{\% pol at 1.3 mm} & \colhead{\% pol at
3.3 mm} }
\startdata

South Cloud & ($-09\arcsec,-11\arcsec$) & $8\arcdeg\pm5\arcdeg$ & $14\arcdeg\pm6\arcdeg$ & $7.3\%\pm1.4\%$ & $5.9\%\pm1.3\%$ \nl
Hot Core & ($-01\arcsec,-03\arcsec$) & $100\arcdeg\pm5\arcdeg$ & $94\arcdeg\pm6\arcdeg$ & $3.8\%\pm0.6\%$ & $2.9\%\pm0.7\%$ \nl
North West (BN) & ($-08\arcsec,+03\arcsec$) & $31\arcdeg\pm9\arcdeg$ & $13\arcdeg\pm7\arcdeg$ & $5.6\%\pm1.8\%$ & $6.6\%\pm1.9\%$ \nl
North Cloud & ($+8\arcsec,+19\arcsec$) & $48\arcdeg\pm8\arcdeg$ & $52\arcdeg\pm8\arcdeg$ & $15.7\%\pm5.3\%$ & $9.7\%\pm3.4\%$ \nl

\enddata

\tablenotetext{a}{Offsets are measured in arcseconds East and North from IRc2}

\end{deluxetable}

\plotone{3mm.eps}

\plotone{1mm.eps}

\end{document}